# Unveiling AC electronic properties at ferroelectric domain walls

*Jan Schultheiß[1,\*], Tadej Rojac[2], and Dennis Meier[1,\*]*

[1] Department of Materials Science and Engineering, Norwegian University of Science and Technology (NTNU), 7034 Trondheim, Norway
[2] Electronic Ceramics Department, Jozef Stefan Institute, 1000 Ljubljana, Slovenia
\* corresponding authors: jan.schultheiss@ntnu.no; dennis.meier@ntnu.no

**Ferroelectric domain walls exhibit a range of interesting electrical properties and are now widely recognized as functional two-dimensional systems for the development of next-generation nanoelectronics. A major achievement in the field was the development of a fundamental framework that explains the emergence of enhanced electronic direct-current (DC) conduction at the domain walls. In this Review, we discuss the much less explored behavior of ferroelectric domain walls under applied alternating-current (AC) voltages. We provide an overview of the recent advances in the nanoscale characterization that allow for resolving the dynamic responses of individual domain walls to AC fields. In addition, different examples are presented, showing the unusual AC electronic properties that arise at neutral and charged domain walls in the kilo- to gigahertz regime. We conclude with a discussion about the future direction of the field and novel application opportunities, expanding domain-wall based nanoelectronics into the realm of AC technologies.**

## 1. Introduction

Ferroelectric domain walls are natural interfaces with a structural width that can approach the unit cell length scale. The domain walls separate volumes with different orientation of the ferroelectric order parameter (**Figure 1**a-c) and can develop physical properties that are completely different from the surrounding domains.[1-6] Although it has been clear for decades that domain walls co-determine the macroscopic response of ferroelectrics,[7-9] the walls themselves moved into focus only recently. The domain-wall related research efforts provided new insight into the structural and electronic transport characteristics of the walls[10] and highlighted their potential for applications in nanoelectronics[3, 5, 11, 12]. Following conventional approaches used in silicon industry, first studies demonstrated the engineering of the electrical conductivity at domain walls via chemical doping,[13-20] annealing in controlled oxygen atmospheres[21-28] or, alternatively, by utilizing the polarization charge as reconfigurable quasi-dopant[29]. An important degree of freedom that is unique to the domain walls and not available at conventional functional interfaces is their spatial mobility. The domain walls readily respond to external stimuli, such as electrical and mechanical fields, enabling real-time adjustment of their local position, electronic configuration, and density.[30, 31] Based on the emergent physical properties at ferroelectric domain walls, first test devices have been realized and their basic functionality has been demonstrated in proof-of-concept studies, including non-volatile memory,[11, 32-34] switches,[35-37] diodes,[38] electronic power conversion,[22, 36] and memristors.[39, 40] Going beyond the unusual electronic response at domain walls, their dielectric,[41-43] piezoelectric,[42, 44, 45] and mechanical properties[46] have drawn attention. In addition, local photovoltaic effects,[47-50] magnetoresistive properties,[51-53] static negative capacitance,[54] and the possibility to control the propagation of phonons and heat flux[55-57] have been investigated. For a more detailed overview of the different application opportunities for domain walls, we refer the reader to recent reviews that address the progress in the field of domain wall nanoelectronics.[1-6, 12]

In this review, we focus on the electrical response of ferroelectric domain walls under alternating voltages (AC). We provide an overview of selected experimental techniques and key results gained in the kilo- to gigahertz regime (1 kHz – 1 GHz). A basic introduction into the electronic transport behavior at ferroelectric domain walls under direct voltages (DC) and alternating voltages is given in section 1. Section 2 addresses the low-frequency regime ($f \lesssim$ 1 MHz), reviewing applied experimental techniques and results. The high-frequency regime ($f \gtrsim$ 1 MHz) is addressed in Section 3. Going beyond the response of individual walls, we discuss how their local properties can be used to tailor the macroscopic behavior of ferroelectric bulk materials (Section 4), closing with a perspective on future challenges and opportunities (Section 5).

### 1.1. DC conduction at ferroelectric domain walls

In 1973, Vul *et al.* estimated the density of free charge carriers that arises at charged ferroelectric domain walls and proposed to use such walls for controlling electrical currents.[58] Around the same time, the existence of conducting domain walls in various ferroelectrics was discussed, including, e.g., $Mg_3B_7O_{13}Cl$,[59] $C_6H_{17}N_3O_{10}S$,[60] $Pb_5Ge_3O_{11}$,[61] $PbTiO_3$,[62] $BaTiO_3$,[63] and $LiNbO_3$.[64] First direct evidence for the emergence of enhanced electronic conductivity at ferroelectric domain walls was provided in 2009. Using conductive Atomic Force Microscopy (cAFM), Seidel *et al.* revealed enhanced electrical conductance at ferroelectric 180° and 109° domain walls in an otherwise insulating $BiFeO_3$ thin film.[65] Today, such cAFM measurements have evolved into a mainstream technique for characterizing the DC transport behavior at ferroelectric domain walls. In cAFM, an electrically biased conducting tip with a diameter of about 5–100 nm is scanned line-by-line over the surface in contact mode, measuring the local conductance with nanoscale spatial resolution. Using this experimental approach, domain walls with enhanced conductance were demonstrated in various ferroelectric materials, including proper ferroelectrics (e.g., $BiFeO_3$,[18, 65-70] $BaTiO_3$,[71, 72] $Pb(Zr,Ti)O_3$,[73] $(K,Na)NbO_3$,[74] and $LiNbO_3$[75-77]) as well as improper ferroelectrics (e.g. $(Ca,Sr)_3Ti_2O_7$,[78] $Cu_3B_7O_{13}Cl$,[79] and hexagonal (h-) $RMnO_3$ ($R$ = Sc, Y, In or Dy–Lu)[80, 81]).

A major achievement in the field was the development of a fundamental framework that allows for explaining the emergence of enhanced electronic conduction. As discussed by Nataf *et al.* in ref. [6], the three main origins of conducting walls are: (i) Differences in the electronic structure at domain walls may locally reduce the band gap. (ii) Domain wall bound charges can shift the valence and conduction bands. (iii) Migration of charged point defects, such as electrons or holes, to the walls can enhance the charge carrier density at the wall and thus its conductivity. Readers interested in a more comprehensive coverage of materials with conducting domain walls and the fundamental mechanisms leading to enhanced electronic DC conduction are referred to general reviews.[4-6]

### 1.2. AC response at domain walls

In contrast to the DC conduction at ferroelectric domain walls, their AC responses and application opportunities in AC circuitry are much less explored. At the bulk level, impedance (dielectric) spectroscopy is a well-established tool,[82, 83] which has been applied extensively to study the macroscopic electronic properties of different solid state systems, including electroceramics[84] and ferroelectrics[85]. This method, supported by modelling, allows for distinguishing between capacitive and resistive responses, as well as contact contributions.[84, 86, 87] In order to benefit from this advantage in local measurements and map the frequency-dependent electronic properties of ferroelectric domain walls, different nanoscale spectroscopy methods have been suggested as we will discuss in Section 2 and 3.

A general overview of the materials that were investigated with respect to the AC properties at ferroelectric walls is given in Figure 1d, listing the applied experimental methods and respective frequency regimes. The frequency-dependent response at domain walls in LiNbO$_3$[88] and h-ErMnO$_3$,[22, 36] was investigated in the kilo- to megahertz regime. One important finding of these studies was that electrode–wall junctions can be used to rectify electrical currents, exhibiting distinct diode-like properties at low frequencies that distinguish them from the domains (see Section 2.6 for details).[22, 36] In the gigahertz regime, a substantial enhancement of the AC conductivity at ferroelectric domain walls and phase boundaries was observed in different host materials, namely BiFeO$_3$,[89] Pb(Zr,Ti)O$_3$,[90] LiNbO$_3$,[91] KNbO$_3$,[92] h-(Lu,Sc)FeO$_3$,[93] and h-YMnO$_3$.[94] At domain walls in BiFeO$_3$,[89] and Pb(Zr,Ti)O$_3$,[90] for example, this increase was related to hopping/tunneling of charge carriers or domain wall vibrational modes, respectively. A more detailed discussion about the different studies and the key findings will be presented in Section 3.2, summarizing the state of the art and the current understanding of the complex physical mechanisms that give rise to the unusual AC responses.

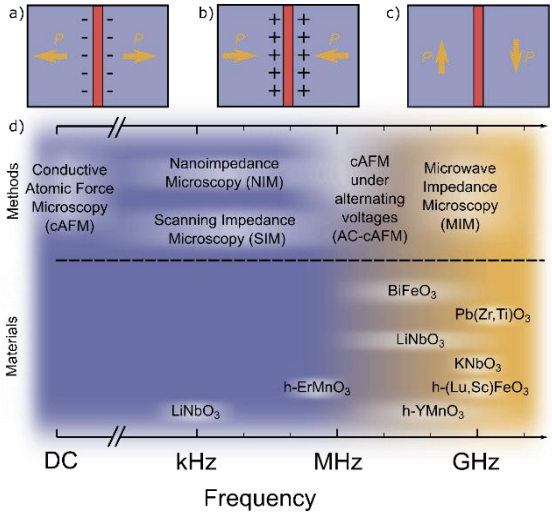

**Figure 1.** Schematic illustration of charged and neutral ferroelectric domain wall configurations and overview of the experimental methods and ferroelectric materials that have been investigated in the different frequency regimes. Depending on the orientation of the spontaneous polarization, $P$ (indicated by yellow arrows), with respect to the domain wall (displayed in red), a) negative, b) positive, and c) neutral charge states may occur. Positive (+) or negative (−) bound charges are indicated.[95, 96] d) The DC conduction at ferroelectric domain walls is commonly characterized using conductive Atomic Force Microscopy (cAFM), whereas Nanoimpedance Microscopy (NIM),[97, 98] Scanning Impedance Microscopy (SIM),[99] and cAFM under AC-drive voltage (AC-cAFM),[22] have been applied to investigate their AC response for frequencies $f \lesssim 1$ MHz. For higher frequencies, Microwave Impedance Microscopy (MIM)[100] is an established method for impedance measurements at the local scale (LiNbO$_3$,[88, 91] h-ErMnO$_3$,[22, 36] h-YMnO$_3$,[94] (Lu,Sc)FeO$_3$,[93] KNbO$_3$,[92] Pb(Zr,Ti)O$_3$,[90] and BiFeO$_3$[89]). The background color indicates two regimes corresponding to kilo- to megahertz ($f \lesssim 1$ MHz, blue) and gigahertz ($f \gtrsim 1$ MHz, yellow) frequency regimes, where domain wall responses are dominated by different contributions as discussed in sections 2 and 3, respectively.

### 1.3. The equivalent circuit model for local AC measurements

The AC behavior at ferroelectric domain walls can be analyzed using an equivalent circuit model analogous to macroscopic impedance spectroscopy measurements. The goal of this section is to introduce the basic framework that is used to rationalize the AC responses obtained in local transport measurements and illustrate the fundamental challenges that arise in local probe experiments.

The electronic response of a material to an electrical stimulus is commonly referred to as electrical impedance $Z$; its reciprocal is the electrical admittance $Y$. To properly analyze local AC measurements, the individual contributions to the impedance, also referred to as electrical microstructure,[84, 85] need to be disentangled. State-of-the art systems that allow for obtaining spatially resolved impedance maps at the nanoscale are typically based on Atomic Force Microscopy (AFM). Comprehensive discussions on the particular aspects of AFM can be found, for example, in reviews by Kalinin et al.,[101] Soergel,[102] and Gruvermann et al.[103]. The typical electrode configuration in AFM is sketched in **Figure 2**.[104] The two main contributions in local impedance measurements are highlighted by the dashed boxes, originating from the tip-sample system (Figure 2a) and parasitic stray effects (Figure 2b). These contributions are discussed in the following.

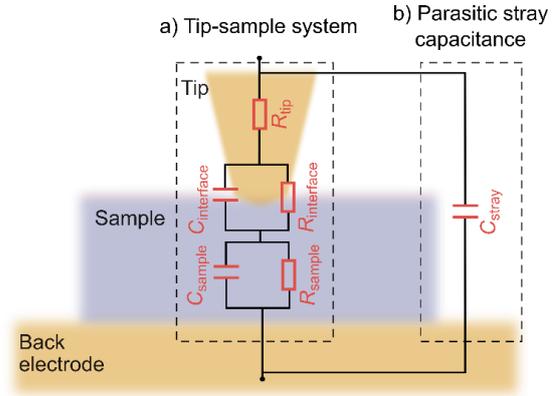

**Figure 2.** Equivalent circuit model for the tip-sample system in a standard AFM setup. The measured impedance has two main contributions, corresponding to the (a) tip-sample system and (b) parasitic stray capacitance. a) The individual contributions to the tip-sample system are represented as circuit elements. The resistance of the tip is represented by $R_{tip}$. The electrical behavior of the tip-sample interface is described by a capacitance ($C_{interface}$) and a resistance ($R_{interface}$) connected in parallel. The intrinsic properties of the sample are represented by a capacitor ($C_{sample}$) and a resistor ($R_{sample}$) connected in parallel. b) The parasitic stray capacitance, $C_{stray}$, is connected in parallel to the tip-sample system and originates from non-localized contributions from the tip-cone, cantilever, and sample holder.

#### 1.3.1. The tip-sample system

The impedance contributions from the tip, interface, and sample are illustrated in Figure 2a, where they are represented by equivalent circuit elements. In the equivalent circuit in Figure 2a, the AFM tip is described by a single resistor with resistance $R_{tip}$. Typical resistivities for commercially available conductive tips are in the range of $3\cdot10^{-5}$–$2\cdot10^{-4}$ Ωm, resulting in $R_{tip} < 10$ kΩ if measured on a highly conducting silver or platinum surface (tip radius ≈ 100 nm).[98]

The interface between tip and sample can be described by a leaky capacitor, modeled as a $RC$ circuit with a resistor ($R_{interface}$) and a capacitor ($C_{interface}$) in parallel. As ferroelectric materials often exhibit semiconducting properties, a Schottky-like barrier with an insulating depletion region forms between the tip and the sample surface.[98, 104, 105] At the interface between Pb(Zr,Ti)O$_3$ and a metallic AFM tip, for example, the interfacial layer was estimated to be 3 to 35 nm thick.[106] Furthermore, the impedance of the $RC$ circuit is a function of the contact force, contact area, and the applied electric field.[100, 104] Other interfacial contributions may arise, e.g., from the formation of a dead layer or oxidation of the surface-near region.[107, 108]

A second circuit component is used to model the intrinsic impedance of the sample, consisting of a capacitor ($C_{sample}$) and a resistor ($R_{sample}$) connected in parallel.[109] The values of $C_{Sample}$ and $R_{Sample}$ are predominantly determined by the sample volume right below the tip. By approximating the probe tip as a circular electrode with diameter $d$, positioned far away from an extended back electrode, one finds that most of the applied voltage drops close to the tip-sample contact.[110, 111] For an electronically homogeneous sample, about 75% of the measured electrical impedance originates from a hemisphere under the tip with a radius of $2d$.[112, 113] Assuming typical tip diameters of $d$=20–200 nm, this corresponds to a hemisphere with a radius of 40–400 nm. It is important to note, however, that ferroelectrics with networks of conducting and/or insulating domain walls are highly inhomogeneous electronic systems,[114] requiring a more case-specific analysis.[115] Furthermore, the electronic properties of the sample (e.g. permittivity[116, 117] and conductivity[118]) will co-determine the distance across which the voltage drops and need to be taken into account to gain quantitative information. Specific examples for the distribution of the electric field under an AFM tip are provided in refs. [119-121].

### 1.3.2. Parasitic stray capacitance

Local parasitic stray capacitances, $C_{stray}$, exist in addition to the contributions from the tip-sample system as illustrated in Figure 2b. Stray capacitances, also termed fringe capacitance,[122] originate from both the upper part of the tip[123, 124] and the cantilever[125] and can be several orders of magnitude larger than the actual local capacitance of the material, $C_{sample}$.[126] Values of $C_{stray}$ =$10^{-10}$ F have been reported, whereas material properties may be as low as $10^{-18}$ F.[98, 127, 128] Parasitic stray effects are a common challenge in local probe measurements leading, for example, to a bias in polarization hysteresis loops in switching experiments.[129, 130] To quantify contributions from the tip-sample system (Figure 2a), the stray capacitance needs to be reduced or, ideally, completely suppressed. One experimental solution is to include a variable RC circuit, compensating for parasitic stray capacitance based on distance-dependent calibration measurements.[122, 131-136] The stray capacitance, however, usually varies also with the lateral position of the probe, e.g., due to tilting of the cantilever, which hinders precise calibration in imaging experiments.[131] To reduce the stray capacitance on the sample side, guard electrodes have been applied.[136, 137] On the tip side, the stray capacitance contributions were reduced by shielding the cantilever. For example, the stray capacitance can already be substantially reduced by inserting a metal foil.[132] The use of electrically shielded cantilevers can further lower parasitic stray contributions.[138-140] Dedicated probes, electrically shielded all the way down to the tip, are commercially available and facilitate reduced stray capacitances down to $10^{-12}$ F (an example is displayed in **Figure 3**).[141-143] Due to the complexity of the problem, however, additional analytical or numerical calculations of the tip-sample interaction are usually necessary to extract accurate quantitative impedance information for the tip-sample system from the experimental data.[117, 118, 144-146]

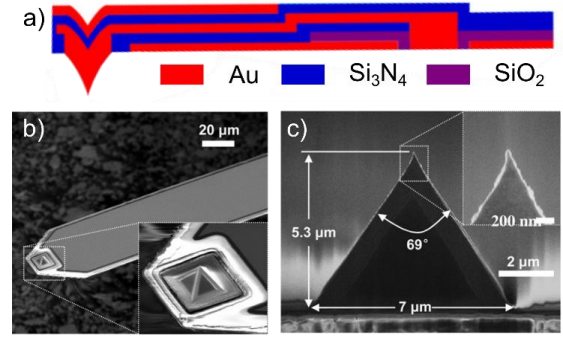

**Figure 3**. Example of a cantilever probe with electrical shielding utilized for nanoscale impedance imaging. a) Schematic of the cantilever profile. The tip and back side of the cantilever is covered by a metal (red) which is electrically grounded. b) Scanning electron microscopy image of the cantilever and the metallic tip. c) Side view of the tip and the sharp tip apex. The diameter of the tip apex is less than 50 nm. Reproduced with permission.[141] Copyright 2012, IOP Publishing.

## 2. Local AC measurements in the kilo- to megahertz regime

We begin this section by reviewing different experimental techniques that have been applied to measure electrical impedance with nanoscale spatial precision at frequencies up to 1 MHz (Figure 1). A schematic illustration of the experiments is provided in **Figure 4**, featuring microelectrodes (Figure 4a), AFM-based approaches, such as NIM, SIM and AC-cAFM (Figure 4b), and microscale bicrystals (Figure 4c). A short introduction to the experimental approaches will be provided along with instructive examples for their application, showing their potential for the study of the local AC response at ferroelectric domain walls. In the second part, we show examples for domain-wall related AC studies, presenting spatially resolved low-frequency data gained on ferroelectric walls in LiNbO$_3$ and h-ErMnO$_3$.

### 2.1. Microelectrodes

An established approach that facilitates electrical impedance measurements at the local scale is to utilize patterned electrodes (typically 1-250 μm in diameter[147]), which are contacted by sharp needles as illustrated in Figure 4a. This technique was used, for example, to analyze the frequency-dependent transport within grains, at grain boundaries, and dislocations in polycrystalline ceramics.[115, 149-158] Furthermore, it found broad application in electrochemistry,[159, 160] biochemistry,[161] and semiconductor technology[162-164]. A comprehensive review of experimental studies is provided in ref. [147]. A fundamental challenge is the time-consuming printing and contacting of the individual electrodes, which was usually realized via lithography-based techniques. To overcome this limitation, in some experiments sharp needles were directly pressed into the surface of the sample.[165, 166] Independent of the method, however, the spatial resolution is limited by the size of the contact area.[149] Concerning ferroelectric domain walls, microelectrodes have been applied to study, for example, the DC transport at individual domain walls in $BaTiO_3$,[71] domain-wall related magnetoresistive[52] and photovoltaic[50] effects in $BiFeO_3$, and resistance-tuning via magnetic domain wall motion in $Nd_2Ir_2O_7$.[53] Although these studies were restricted to DC currents, they demonstrated the possibility to contact individual domain walls via microelectrodes, opening the door for future frequency-dependent investigations of the local AC response.

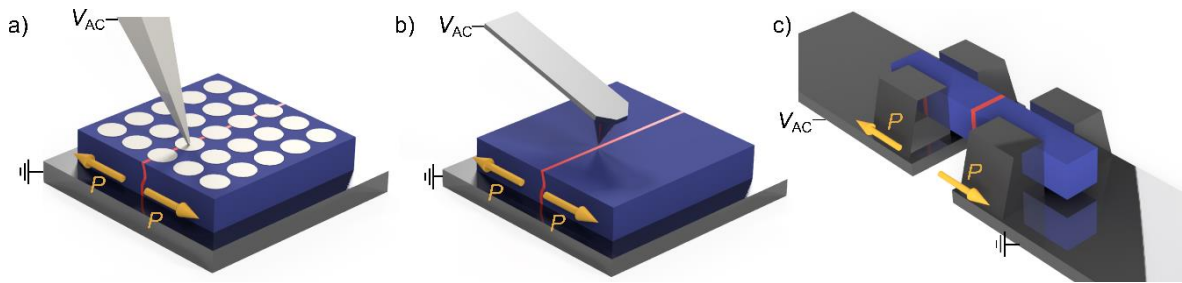

**Figure 4:** Illustration of local characterization techniques that allow for measuring electronic AC transport properties in the kilo- to megahertz regime (f ≲ 1 MHz). a) Frequency-dependent conductivity studies of individual ferroelectric domain walls can be realized using microelectrodes[147] (white disc) placed in domain (blue) and domain wall (red) regions. This approach allows for resolving relative changes in the AC response that arise at the position of the walls (yellow arrows indicate the polarization direction within the domains). b) The application of AFM-based techniques, such as NIM,[97, 98] SIM,[99] and AC-cAFM,[22] enables spectroscopy measurements at ferroelectric domain walls with nanoscale spatial resolution. c) Analogous to bicrystals,[148] which were used in studies on grain-boundary-related effects in polycrystalline ceramics, the investigation of nanostructured ferroelectrics with only one domain wall represents a powerful approach for future frequency-dependent investigations.

2.2. Nanoimpedance Microscopy

Scanning probe techniques have successfully been applied to realize local electrical impedance measurements since the 1980s.[167, 168] In the original experiments, a sharp probe was scanned over the surface of metallic alloys or materials with organic coatings, achieving a spatial resolution of about 100 µm. With the progress in AFM, the resolution was pushed towards the nanoscale.[169-172] A breakthrough in spatial resolution was achieved in 2003, realizing local impedance spectroscopy measurements with a conducting AFM tip, facilitating a contact area with a radius of ~100 nm (Figure 4b).[97, 98] This AFM-based approach is commonly referred to as NIM, where the AFM tip serves as a spatially mobile electrode.[173-175] A constant DC bias voltage and a superimposed probing AC voltage of variable frequency are applied to the back electrode of the sample. The constant DC bias is necessary to access intrinsic properties at $f \lesssim 1$ MHz,[128] whereas the electrical impedance is measured at the same frequency as the applied driving field frequency. It is important to note, however, that while a small contact area permits high resolution, it also leads to high impedance values of the tip-sample system. Values for the measured spreading resistance and capacitance, $R_{spr}$ and $C_{spr}$, can be estimated by representing the AFM tip by a circular electrode of diameter $d$:[112, 113, 176]

$$R_{spr} = \frac{1}{2d\sigma_{sample}} \quad \text{Equation 1}$$

$$C_{spr} = 2d\varepsilon_{sample} \quad \text{Equation 2}$$

In equation 1 and 2, $\sigma_{sample}$ and $\varepsilon_{sample}$ denote the local conductivity and permittivity of the sample, respectively. For values typical for ferroelectric materials, high spreading resistances in the order of $10^9$ to $10^{12}$ Ω are expected, whereas the spreading capacitance is rather low ($\approx 10^{-15}$ to $10^{-18}$ F). Thus, to measure high resistances and low capacitances is one of the main challenges in NIM.[98] Most feasible for detection are lock-in based techniques, which readily facilitate a resistance and capacitance sensitivity of $10^{15}$ Ω and $10^{-18}$ F, respectively, over a broad frequency range from a few Hz to several MHz[122, 132, 135, 177] with the potential to reach zeptofarad ($10^{-21}$ F) resolution in dedicated setups.[127, 178, 179]

NIM is well-established within the electrochemistry community and frequently applied to explore, for example, electrode heterogeneities[180-182] or defects in organic coatings.[183-185] In condensed matters physics, NIM has been utilized to gain access to characterize grain boundaries in polycrystalline boron-doped diamond thin films,[186] silver ion conducting glasses,[128] polycrystalline ZnO varistors,[97, 187] and other electronically inhomogeneous materials.[137, 177, 188-190] Systematic studies over a broad frequency range, which are required to disentangle individual impedance contributions are, however, rather rare. One example is the work by Layson et al., presenting NIM spectroscopy data recorded over a broad frequency range from 400 Hz to 100 MHz on a polymer electrolyte film.[182] Furthermore, Kruempelmann et al. pointed out that AC frequencies above 1 MHz are required to access the intrinsic sample properties (i.e., $R_{sample}$ and $C_{sample}$ in Figure 2a).[128] First NIM experiments on ferroelectric domain walls were performed in $LiNbO_3$ single crystals,[88] revealing enhanced AC conductivity with respect to the domains as discussed in more detail in section 2.6.1.

2.3. Scanning Impedance Microscopy

SIM is an AFM-based approach for local impedance measurements, where the data, unlike NIM, is recorded in non-contact mode. A constant bias voltage is applied to the probe tip and a lateral bias voltage is applied across the sample via macroscopic electrodes.[99, 191, 192] The lateral bias voltage induces an oscillation in the electric surface potential, which triggers a periodic mechanical vibration of the cantilever via the electric far field. The corresponding amplitude and phase of the mechanical vibration of the cantilever is read out using a lock-in amplifier. In combination with an AFM unit, both parameters can be obtained spatially resolved. As derived in refs. [99, 191], the measured amplitude and phase allow for determining local capacitive and resistive values. SIM was applied, for example, to study the frequency-dependent dielectric response at grain boundaries in $SrTiO_3$ bicrystals,[193] p-doped silicon,[191] carbon nanotubes,[194, 195] as well as grain boundaries in $CaCu_3TiO_4O_{12}$ and $BiFeO_3$ polycrystalline ceramics.[192, 196] In the study on $BiFeO_3$, it was concluded that ferroelectric domain walls do not contribute to measured signals at kHz frequencies in the investigated samples.[192] However, to understand how the unusual DC conductivity of ferroelectric domain walls affects the AC transport properties at low frequencies, more comprehensive measurements on different types of walls are highly desirable, representing an intriguing challenge for future SIM studies.

2.4. cAFM under alternating voltages

Going beyond basic DC measurements, cAFM can be performed under AC voltages referred to as AC-cAFM. In AC-cAFM, an AC voltage is applied to the sample via the tip or the back electrode and the local electric response is mapped while scanning the tip over the surface in contact mode.[22, 36] Unlike in SIM and NIM, the high-frequency components are filtered out via a low pass filter, so that the measured signal represents the AC-driven DC component, i.e., the 0 Hz component. This DC component is non-zero whenever the tip-sample system exhibits asymmetric current-voltage characteristics under bipolar drive voltage.[106, 197, 198] By measuring the AC-cAFM signal as function of the frequency and amplitude of the drive voltage, information about the local AC behavior in the kHz to MHz regime can be obtained. The latter is possible as the behavior of the tip-sample contact is co-determined by the material's band structure and charge carrier mobility (a recent review on the physical and chemical aspects of the barrier formation can be found in ref. [199]). The AC-cAFM method was utilized to measure the AC response at neutral[22] and charged[36] domain walls in improper ferroelectric h-$ErMnO_3$ as presented in Section 2.6.2.

2.5. Microscale bicrystals

Impedance measurements of individual microstructural components (i.e., grain boundaries) have been realized by isolating the features of interest in macroscopic bicrystals.[200-202] While individual ferroelectric domain walls have been isolated in, e.g., $Gd_2(MoO_4)_3$ single crystals,[203] the impedance properties have not been studied. Recently, this macroscopic approach was transferred to the nanoscale and a single grain boundary of a ceria fiber was extracted using focused ion beam (FIB). This facilitated impedance characterization of the grain boundary from 1 mHz to 1 MHz at temperatures from 300°C to 500°C.[148] Although FIB-based extraction of domain walls in ferroelectric materials has been demonstrated,[114, 204] the impedance characteristics of the microscopic crystals as proposed in Figure 4c have so far not been analyzed.

### 2.6. NIM- and AC-cAFM-based studies on ferroelectric domain walls

#### 2.6.1. Enhanced AC domain wall response in LiNbO$_3$

Charged head-to-head domain walls in LiNbO$_3$ exhibit enhanced electronic DC conduction,[75-77, 95] reaching values up to 13 orders of magnitude larger than in the surrounding domains.[205] The physical properties of the walls have been studied intensively and innovative domain-wall based device concepts have been proposed.[34, 37, 40] In contrast, much less is known about the AC behavior of the conducting head-to-head walls in LiNbO$_3$. In pioneering NIM experiments, the AC response of these walls was measured as reported in ref. [88]. The impedance characteristics of the positively charged head-to-head domain walls are presented in the spatially resolved measurements in **Figure 5**a. The maps of the real and imaginary part of the AC current show a clear contrast between the domain walls and the domains, indicating an enhanced AC conductivity at the domain walls for 100 Hz, which was related to charge carrier hopping. To obtain the spatially resolved impedance images, experimental parameters, such as frequency and the amplitude of the AC signal, were tuned and a DC offset signal was applied to optimize the domain wall contrast. A frequency of 800 Hz was determined as the upper limit for domain wall contrast based on the NIM experiments. For higher frequencies, the wall contrast vanished, which was related to a dominating domain response and/or the impact of parasitic stray capacitance contributions (see Figure 2a). Finally, while the authors consider the lossy contribution related to domain wall displacements under applied AC fields weak, only a full analysis considering both the conductive and dynamic domain wall contributions can reveal the complex AC behavior of domain walls (see also section 3.2).

The study on charged head-to-head domain walls in LiNbO$_3$[88] represented an important step regarding domain wall imaging in the low-frequency AC regime, as it demonstrated the general feasibility and potential of local NIM measurements. The application of shielded tips (see Figure 3) may further improve the information extraction and guided by theoretical analysis of the NIM spectra, facilitate new insight into the physics of ferroelectric domain walls under AC drive voltages. Particularly interested could be, for example, to characterize the AC behavior of domain walls with different concentration of accumulated charge carriers.[13, 20]

#### 2.6.2. Electrical half-wave rectification at domain walls in h-ErMnO$_3$

Hexagonal manganites exhibit geometrically driven improper ferroelectricity.[206] The material family is intriguing as model system for domain wall studies as it naturally forms all possible charge states of 180° ferroelectric domain walls in the as-grown state, i.e., neutral side-by-side, positively charged head-to-head, and negatively charged tail-to-tail walls (Figure 1a-c) as discussed in ref. [80]. Local measurements (cAFM) on the p-type semiconductor h-ErMnO$_3$ revealed that hole carriers accumulate at tail-to-tail walls, whereas their density is reduced at head-to-head walls, resulting in enhanced and reduced DC conductance, respectively. In addition, neutral domain walls have been reported to exhibit enhanced DC conductance[22, 207] with secondary contributions from charged wall sections below the surface.[23, 208]

The AC response at the different types of ferroelectric 180° domain walls in h-ErMnO$_3$ was investigated by AC-cAFM as shown in Figure 5b-d. Figure 5b presents an AC-cAFM scan recorded at 1 MHz on a single crystal with out-of-plane polarization, showing pronounced contrast at the neutral domain walls. The wall contrast was observed to vanish at ~1-10 MHz, indicating a frequency-dependent transition from asymmetric to symmetric current-voltage characteristics. Analogous to macroscopic spectroscopy measurements, the respective cut-off frequency, $f_c$, correlates with the local conductivity. Thus, as the cut-off frequency at the walls was higher than for the domains, it was concluded that the neutral domain walls exhibit enhanced electronic conduction under the applied AC voltage.[22] The enhanced conduction was explained based on oxygen interstitials, which accumulate at the neutral walls due to a lower formation energy compared to the bulk, increasing the density of mobile hole carriers.[22, 23] AC-cAFM data obtained from charged ferroelectric domain walls in h-ErMnO$_3$ is displayed in Figure 5c. The AC-cAFM scan was taken at 0.5 MHz and shows suppressed and enhanced AC-driven current signals at head-to-head and tail-to-tail domain walls, respectively. Systematic frequency-dependent spectroscopy measurements in the frequency range 0.1-10 MHz (Figure 5d) revealed distinct cut-off frequencies, $f_c$, for the two types of domain walls with $f_c^{\rightarrow\leftarrow} < f_c^{\text{Domain}} < f_c^{\leftarrow\rightarrow}$, consistent with the transport behavior observed in DC conductance measurements, $\sigma$ ($\sigma^{\rightarrow\leftarrow} < \sigma^{\text{Domain}} < \sigma^{\leftarrow\rightarrow}$.)[80]

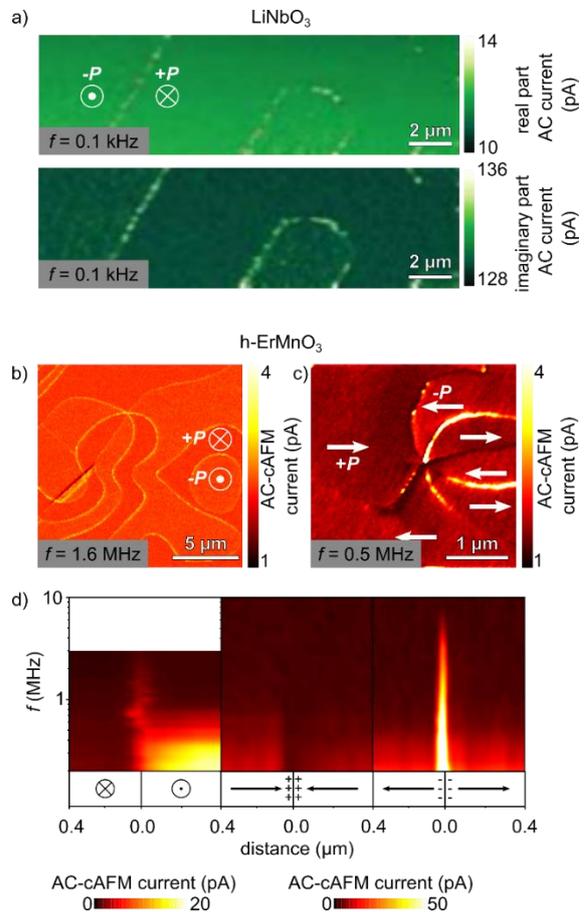

**Figure 5.** Spatially resolved measurements of the unusual AC response at different ferroelectric domain walls in the kilo- to megahertz regime. a) NIM scans showing the real and the imaginary part of the AC current response measured at 0.1 kHz in LiNbO$_3$. Reproduced with permission.[88] Copyright 2014, IOP Publishing. b) AC-cAFM map acquired on h-ErMnO$_3$ (out-of-plane polarization) at 1.6 MHz. Pronounced contrast is observed at the neutral domain walls. Reproduced with permission.[22] Copyright 2018, Springer. c) AC-cAFM image (0.5 MHz) recorded on h-ErMnO$_3$ with in-plane polarization, revealing enhanced and suppressed AC-cAFM current contrast at tail-to-tail and head-to-head domain walls, respectively. d) Spectroscopic measurements of the AC-cAFM current at different

types of domain walls in h-ErMnO$_3$, recorded from 0.1-10 MHz. Reproduced with permission.[36]

The frequency-dependent data gained on ferroelectric domain walls in h-ErMnO$_3$ were rationalized based on a simplified version of the equivalent circuit model in Figure 2a, considering $R_{sample}$ and a voltage-dependent $R_{interface}$. The results are intriguing as they demonstrate the possibility to control the local AC response in the kilo- to megahertz regime via multiple parameters, including the domain-wall charge state,[36] defect concentration,[22] and the applied drive-voltage amplitude[36]. Enabled by this tunability, it has been proposed to develop ultra-small half-wave rectifiers based on ferroelectric domain walls, facilitating AC-to-DC signal conversion at the nanoscale.[22, 36]

In conclusion, at low frequencies the electrode-wall contacts are playing a key role, offering intriguing possibilities for the development of nanoscale hetero-junctions for AC applications, where the active area is defined by the width of the domain wall. At present, however, investigations are still limited to a few proof-of-concept experiments and more comprehensive studies are highly desirable, testing different types of electrodes and model systems to explore the opportunities for future AC nanotechnology.

## 3. Probing ferroelectric domain walls at gigahertz frequencies

### 3.1. Microwave Impedance Microscopy

An established near-field technique that facilitates mapping of the dielectric and conductive properties in the gigahertz regime (Figure 1) is MIM.[209-211] MIM is realized by sending a radio or microwave signal (1 MHz - 10 GHz) to a probe, which is brought near or in contact with the surface. By detecting the reflected signal while scanning across the surface, the real (MIM-Re) and imaginary (MIM-Im) part of the tip-sample admittance, i.e., the reciprocal of the impedance, are mapped with high spatial resolution.[212, 213] The electromagnetic field of the reflected radio-/microwaves interacts with a small volume under the probe. Thus, the resolution is mainly determined by the local electric field distribution, which is analytically derived in ref. [117]. Typically, a spatial resolution comparable to the tip diameter is achieved in MIM (see also section 1.3.1),[100] varying with the sample's permittivity[116, 117] and resistivity[118]. In the original experiments, scanning was realized using antennas as probes, which facilitated a spatial resolution of ≈ 100 μm.[214, 215] A substantial leap ahead was achieved by utilizing shielded AFM probes (see Figure 3 and section 1.3.2), which improved the spatial resolution by several orders of magnitude.[116, 139, 216, 217] Today, state-of-the-art AFM-based MIM enables sub-100 nm spatial resolution.[141, 218] Importantly, and in contrast to the experimental methods introduced in section 2, MIM utilizes the interaction between microwaves and the sample, which implies that no additional electrodes or specific contacts are required to access the intrinsic material properties. This degree of freedom makes MIM a highly versatile tool for nanoscale impedance measurements.

Like NIM (Section 2.2), MIM consists of a probe, detection electronics, and the scanning platform. However, instead of lock-in amplifiers, MIM-specific electronics are used that allow operation at microwave frequencies. Efficient power transfer from the 50 Ω transmission lines to the AFM probe is achieved by an impedance match section,[219] as discussed in ref. [94]. For further technical details, the interested reader is referred to, for example, refs. [109, 118, 219]. The measured real (MIM-Re) and imaginary (MIM-Im) channels are proportional to the real and imaginary part of the tip-sample admittance, respectively.[109, 220] To gain quantitative insight and correlate the measured signals to the physical properties of the investigated material (i.e., $R_{Sample}$ and $C_{Sample}$, Figure 2a), additional calculations are required modelling the tip-sample interaction. For specific geometries, analytical solutions can be obtained[117, 221] or, alternatively, numerical simulations are applied to describe the actual tip-sample system.[219, 222] An example for the numerical analysis of the tip-sample interaction is displayed in **Figure 6**, considering the case of a 10 nm thin film, which is located 30 nm under the surface.[118] The calculated quasi-static potentials in Figure 6 are obtained by finite element analysis (FEA) and allow for correlating the resistivity of the thin film, $\rho$, to the MIM-Im and MIM-Re signals.[109, 220]

As a powerful method for high-resolution conductivity and capacitance studies, MIM is widely applied for electrical characterization at the nanoscale, including materials for semiconductor technology,[223] field-effect transistors,[224] as well as strongly correlated electron systems.[225] Readers interested in a more comprehensive coverage are referred to ref. [100]. In the next section 3.2, we provide an overview of recent studies, which applied MIM to investigate the physical AC properties of ferroelectric domain walls at frequencies in the 1 MHz-10 GHz regime.

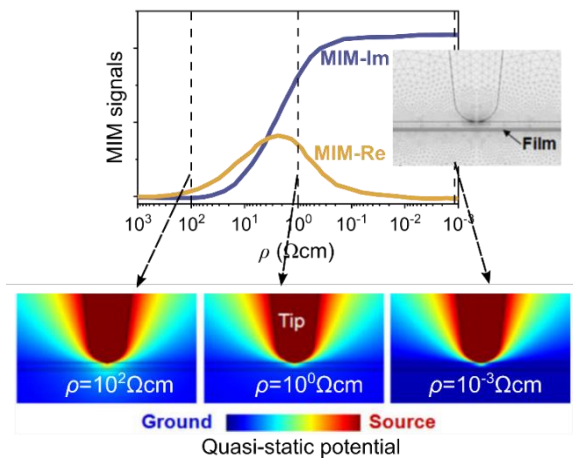

**Figure 6.** Relation between experimental MIM results (MIM-Im and MIM-Re channels) and sample resistivity, $\rho$. To model the tip-sample admittance, FEA is utilized with the dense mesh geometry displaying the tip in the inset. The sample consists of a thin (10 nm) film buried 30 nm below the surface. A typical MIM response curve with the sample resistivity, $\rho$, (Figure 2a) as tuning parameter is simulated. For highly resistive films ($\rho > 10^2 \Omega cm$), the response is dielectric and both channels are low. For highly conductive films ($\rho < 10^{-3} \Omega cm$), the MIM-Im channel gives a large signal, whereas the MIM-Re channel is low. Reproduced with permission.[118] Copyright 2011, Springer.

### 3.2. MIM measurements at ferroelectric domain walls

The advancement of AFM-based MIM experiments enabled nanoscale conductivity measurements at domain walls in ferroelectric thin film and bulk materials. Following the pioneering MIM studies on ferroelectric domain walls by Tselev et al. in 2016,[90] the method was applied to study domain wall responses in various ferroelectric materials as summarized in **Figure 7**. Different microscopic mechanisms have been discussed, explaining the enhanced AC currents at ferroelectric domain walls in comparison to the surrounding domains. Furthermore, innovative applications in low-loss microwave devices[226] and high-frequency waveguides[227] have been proposed. In the following, an overview of recent MIM studies on ferroelectric domain walls is presented with a short discussion of the key observations.

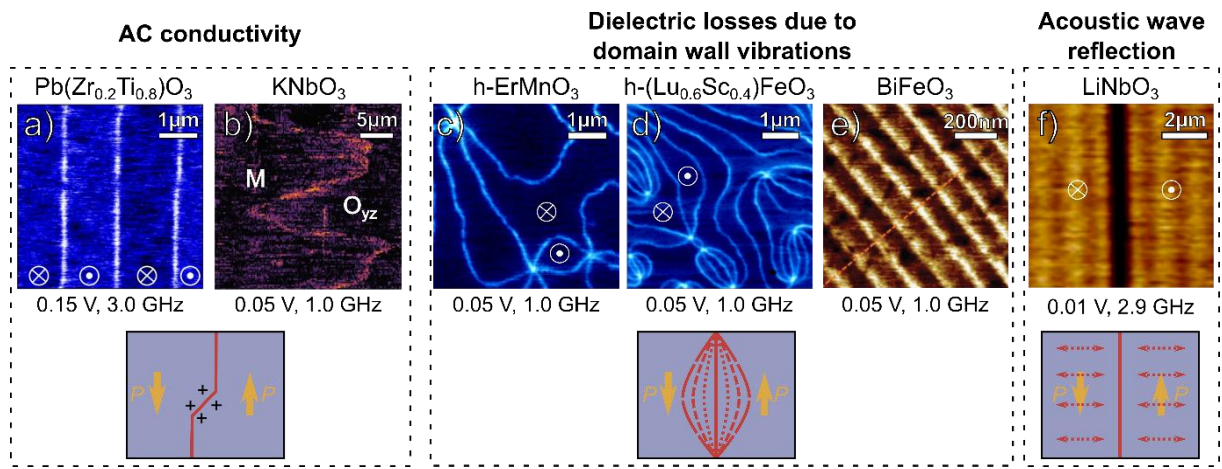

**Figure 7.** Response of ferroelectric domain walls to microwave frequencies. The applied voltage amplitude and certain frequency is indicated. The response is displayed spatially resolved for a) Pb(Zr$_{0.2}$Ti$_{0.8}$)O$_3$ (Reproduced with permission.[90] Copyright 2016, Springer.), b) KNbO$_3$ (Reproduced with permission.[92] Copyright 2017, Wiley-VCH.), c) h-ErMnO$_3$ (Reproduced with permission.[94] Copyright 2017, American Association for the Advancement of Science.), d) (Lu$_{0.6}$Sc$_{0.4}$)FeO$_3$ (Reproduced with permission.[93] Copyright 2018, American Physical Society.), e) BiFeO$_3$ (Reproduced with permission.[89] Copyright 2020, Wiley-VCH.), and f) LiNbO$_3$ (Reproduced with permission.[91] Copyright 2018, National Academy of Sciences.). The abbreviations M and O$_{yz}$ in b) indicate a monoclinic and orthorhombic subphase, respectively. The out-of-plane orientation of the ferroelectric polarization is displayed in a), c), d), and e). The domain walls in e) correspond to 71° domain walls. Yellow arrows in the schematics indicate the orientation of the spontaneous polarization vector and the red solid lines represent the domain wall. The "+" below panels a) and b) denote domain wall bound charges. The dashed and dotted lines below panel c)-e) indicate domain wall vibrations, whereas dashed arrows below panel f) indicate acoustic waves in the domains.

### 3.2.1. AC conductivity at domain walls in Pb(Zr$_{0.2}$Ti$_{0.8}$)O$_3$ and KNbO$_3$

Pb(Zr$_{0.2}$Ti$_{0.8}$)O$_3$ – In their original work, Tselev and co-workers studied the AC conductivity of nominally neutral walls in a Pb(Zr$_{0.2}$Ti$_{0.8}$)O$_3$ thin film (MIM-Re channel, Figure 7a).[90] It was shown that the AC conductivity at 3 GHz was two orders of magnitude higher than the DC conductivity recorded with the same probing voltage. The enhanced AC conduction was attributed to defect-driven domain wall roughening,[228, 229] which forces the walls to tilt away from their ideal charge neutral position, consistent with previous transmission electron microscopy studies.[230, 231] As a consequence, local sections with bound charges arise, which are compensated by mobile carriers as illustrated in Figure 7 (below panels a and b). These charge carriers are localized due to energy barriers between which they can oscillate at GHz frequencies,[232-234] giving rise to the enhanced AC conductance at the wall.

KNbO$_3$ – The AC conductivity of boundaries separating monoclinic (M) and orthorhombic (O$_{yz}$) subphases in KNbO$_3$ was investigated by Lummen et al..[92] Although the phase boundaries are different from the ferroelectric walls discussed so far, the experiments reflect the potential of MIM for domain wall related investigations. To study the electrical properties of the phase boundaries, MIM scans were performed at a frequency of 1 GHz. The meandering boundaries can be seen in Figure 7b (MIM-Re channel), featuring pronounced contrasts in comparison to the surrounding subphases. Complementary scanning X-ray diffraction microscopy measurements revealed localized mechanical strain at the M-O$_{yz}$ interface, which was proposed to locally modify the band gap and, hence, give rise to anomalous conduction at the boundary.

### 3.2.2. Dielectric losses due to domain wall vibrations in h-RMnO$_3$, h-(Lu$_{0.6}$Sc$_{0.4}$)FeO$_3$, and BiFeO$_3$

In addition to oscillating charge carriers at the domain walls, the domain walls themselves can vibrate around their equilibrium position.[235] Such vibrations cause microwave energy dissipation[90] and displacement currents[236], which is also detectable by MIM.

h-RMnO$_3$ – The structural dynamics of ferroelectric domain walls in the microwave regime was first studied by Wu et al..[94] The team investigated the response of different hexagonal manganites over a broad range of frequencies between 1 MHz and 10 GHz, including h-ErMnO$_3$, h-YMnO$_3$, and h-HoMnO$_3$. A representative image (MIM-Re channel) gained on h-ErMnO$_3$ (out-of-plane polarization) is displayed in Figure 7c, showing enhanced contrast at neutral domain walls for an AC frequency of 1 GHz. The authors observed that the conductivity of the neutral domain walls was enhanced by five to six orders of magnitude in comparison to the DC conductivity.

In contrast to the conductivity enhancement at the neutral domain walls, charged domain walls in h-HoMnO$_3$ (in-plane polarization) did not show a detectable MIM contrast at 1 GHz. This finding was explained based on the orientation of the local electric field under the AFM tip with respect to the direction of the spontaneous polarization vector. For neutral domain walls, the vertical component of the oscillating electric field is aligned either with the up or down polarized domains, leading to periodic domain wall vibration associated with dielectric losses. For charged domain walls, the local electric field is oriented perpendicular to the polarization direction and no domain vibrations are excited. Thus, it was concluded in h-ErMnO$_3$ the significant enhancement of the MIM contrast at neutral domain walls in the GHz regime (Figure 7c) originates from dielectric losses due to domain wall vibrations rather than the presence of free charge carriers addressed in section 3.2.1.

(Lu$_{0.6}$Sc$_{0.4}$)FeO$_3$ – The behavior of ferroelectric domain walls in (Lu$_{0.6}$Sc$_{0.4}$)FeO$_3$ single crystals in the microwave regime was investigated by Wu et al..[93] The material belongs to the family of hexagonal ferrites, which is isomorphic to the hexagonal manganites.[237] The authors performed MIM measurements at 1 GHz with different bias offset voltages varying from -3 V to +2 V. A representative image (MIM-Im channel) at a bias offset voltage of 0 V is displayed in Figure 7d for a sample with out-of-plane polarization. Similar to h-ErMnO$_3$,[94] pronounced contrast is observed at the neutral 180° domain walls, distinguishing them from the ferroelectric domains. Furthermore, it was observed that the MIM signal of the domains at 1 GHz strongly depends on the bias offset voltage, which was explained by a Schottky band bending.[105] In contrast, the MIM signal of domain walls was nearly independent on the applied bias voltage, corroborating that the dielectric loss due to domain wall vibrations at microwave frequencies is the predominant contribution to the measured MIM signal.[93]

BiFeO$_3$ - Huang et al. investigated the AC response of 71° domain walls in BiFeO$_3$ (Figure 7e). Their MIM measurements (MIM-Re channel) revealed an AC conductivity of 10$^3$ S/m at 1 GHz, which is about 10$^5$ times higher compared to the DC conductivity (10$^{-2}$ S/m) at the same type of wall.[89] The significant enhancement of the apparent domain wall conductivity under microwave frequencies was explained based on dipolar losses that arise due to vibrations of the 71° domain walls under the application of an alternating electric field. Furthermore, atomistic simulations predicted that intrinsic displacement currents from domain wall vibrations of neutral 109° and 71° domain walls in BiFeO$_3$ can be significant in the GHz regime, representing an additional source for locally enhanced losses.[236]

3.2.3 Acoustic wave reflection at domain walls in LiNbO$_3$

LiNbO$_3$ – The response of domain walls in LiNbO$_3$ to an alternating electric field was measured by Zheng et al., covering frequencies from 0.3-6.0 GHz.[91] A representative image (MIM-Re channel) is displayed in Figure 7f. Suppressed MIM contrast was observed at the position of the domain wall, accompanied by a fringe pattern that spreads out into the adjacent domains. In their work, the authors argued that the number of free carriers at a straight domain wall in z-cut LiNbO$_3$ is negligible and excluded dielectric losses due to domain wall vibrations. Instead, the authors concluded that the MIM contrast reduction at the domain walls originates from acoustic wave reflections. In their model, the fringe pattern in the adjacent domains was related to acoustic waves travelling through the sample, induced by the piezoelectric coupling between the domains and the electric field.

In conclusion, different microscopic models have been proposed to explain the AC responses of ferroelectric domain walls at microwave frequencies, reflecting the complexity of the underlying physics. With the ongoing progress in MIM technology and the theoretical analysis of the tip-sample interaction, however, nanoscale experiments over a broad frequency range are becoming more and more accessible, which will propel the understanding and help developing of a comprehensive framework.

## 4. Engineering macroscopic electromechanical properties utilizing functional domain walls

In the previous sections, we have discussed different local probe techniques and experiments with a focus on the AC transport properties of individual ferroelectric domain walls. The research activities provided new insight into the nanoscale physics of domain walls and expanded the field of domain wall nanotechnology into the realm of AC electronics. The better understanding of the emergent electronic phenomena at ferroelectric domain walls, however, has not just propelled new lines of nano-research; it also promoted property engineering at macroscopic length scales by enabling a much more targeted control via the utilization of domain walls with specific electronic properties. Although it has been known for a long time that domain walls co-determine the macroscopic responses of ferroelectrics, traditional approaches mainly focus on tuning macroscopic properties via the domain wall density[72, 238, 239] and mobility[240-243]. The AC responses which were recently found to be substantially different at the ferroelectric domain walls compared to the surrounding domain represent a new degree of freedom that can be used for tailoring the macroscopic physical properties and, hence, achieve completely new functional behaviors under alternating electric fields.

In this section, we will present how insulating and conducting domain walls impact the macroscopic dielectric and piezoelectric response of bulk ferroelectrics, reviewing recent work on the model systems h-ErMnO$_3$ and BiFeO$_3$ as instructive examples. In section 4.1, we discuss how networks of insulating domain walls in h-ErMnO$_3$ single crystals act as barrier layer capacitors[244, 245], and in section 4.2 we will address the influence of conducting domain walls on the frequency-dependent piezoelectric properties of polycrystalline BiFeO$_3$ ceramics.[67] In both cases, the key concept is to exploit domain-wall related Maxwell-Wagner relaxations, i.e., dielectric or piezoelectric enhancements that arise due to the electrical heterogeneity introduced by the insulating and conducting walls.[246-248]

### 4.1. Insulating ferroelectric domain walls as internal barrier layer capacitors

Interfaces can act as parallel-plate capacitors with very small plate distances, leading to high values for the measured dielectric constant.[249] A prominent model system is CaCu$_3$Ti$_3$O$_{12}$, where "giant" dielectric properties have been reported for polycrystalline ceramics, originating from an electrical heterogeneity caused by semiconducting grains and insulating grain boundaries.[250, 251]

Ruff et al. demonstrated that instead of insulating grain boundaries, ferroelectric domain-wall networks in single crystals can be used to control the AC behavior at the macroscale.[252] The frequency-dependent dielectric response measured on a h-ErMnO$_3$ single crystal at different temperatures is displayed in **Figure 8**a.[244] The data shows two relaxation steps labelled ① and ②, corresponding to a surface barrier layer capacitor (SBLC) and most-likely internal barrier layer capacitor (IBLC), respectively, whereas contribution ③ was attributed to the bulk.[244] Based on systematic dielectric spectroscopy measurements, the authors concluded that the IBLC effect originates from insulating head-to-head domain walls, which act as internal plate capacitors.[244]

The IBLC originating from insulation domain wall is intriguing as the volume fraction of insulating domain walls can readily be controlled using, e.g., chemical doping,[253] electric fields[245] or annealing[254]. The latter two control strategies are applicable even after the material has been synthesized, representing a new degree of freedom for the design of high-κ materials. At present, however, the observation of "giant" dielectric constants due to insulating ferroelectric domain walls is restricted to the family of hexagonal manganites. To further explore the opportunities and limitations associated with the application of ferroelectric domain walls as barrier layer capacitors, other systems that develop insulating domain walls, such as (Ca,Sr)$_3$Ti$_2$O$_7$[78] and Pb(Zr,Ti)O$_3$,[21] may be studied, as well as collective phenomena combining the IBLC effects observed at insulating domain walls and grain boundaries in ferroelectric ceramics.

### 4.2. Tuning the AC piezoelectric response via conducting domain walls

The impact of conducting ferroelectric domain walls on the macroscopic piezoelectric AC response was investigated by performing systematic studies on polycrystalline BiFeO$_3$ as presented in **Figure 9**. Frequency-dependent data from the work is shown in Figure 9a,[20, 67, 255, 256] featuring a low-frequency regime ($f < \sim 10$ Hz) with enhanced piezoelectric properties. It was found that the piezoelectric coefficient in the low-frequency regime shows a strong electric-field dependency (see arrow in Fig. 9a). Alongside the field dependence of the piezoelectric phase angle,[67] the data revealed that the field induced low-frequency piezoelectric enhancement arises from irreversible displacements of non-180° domain walls. This contribution was confirmed by deconvoluting the lattice strain and strain from domain wall motion using in situ XRD diffraction analysis.[256] The unusual low-frequency behavior was explained based on the nonlinear piezoelectric Maxwell-Wagner effect where conductivity at the walls plays a key role.[20, 67, 256] Although significant anisotropy in the electric conductivity of the perovskite BiFeO$_3$ is not expected, presence of conducting domain walls in ceramics, as seen in the cAFM image in the inset to Figure 9a, can introduce electrical inhomogeneities, which is explained next.

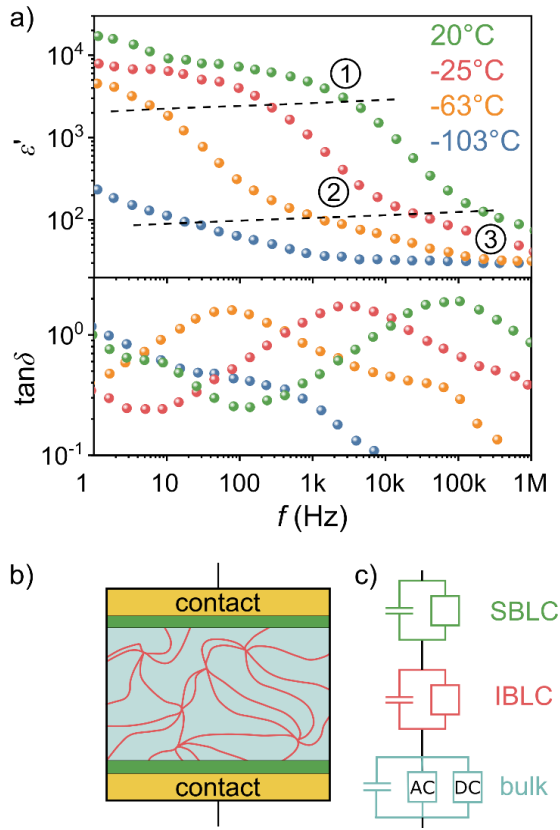

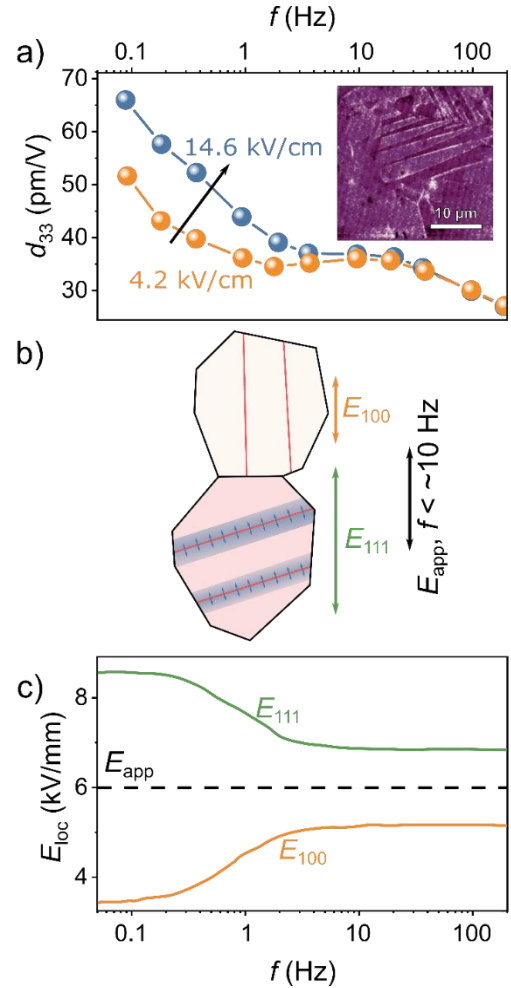

**Figure 8.** Insulating domain walls as barrier layer capacitors. a) Dielectric spectroscopy measurements on an h-ErMnO$_3$ single crystal gained at different temperatures reveal two step-like features in the dielectric constant ($\varepsilon'$), ① and ②, which were related to SBLC and IBLC effects, respectively.[244] The bulk properties, ③, are accessed at high frequencies ($f$ > 0.1-10 MHz). b) Schematic illustration of the different contributions and c) their representation by an equivalent circuit model.[252] A detailed description of the equivalent circuit model can be found in ref. [252]. Reproduced with permission.[252] Copyright 2017, American Physical Society.

Depending on the orientation of the conducting domain walls within a grain, the grain conductivity can vary as schematically depicted in Figure 9b for the simplified case of [100] and [111] oriented grains (with respect to the electric field axis). On the grain level, the effect was modelled analytically by Liu *et al.* for the specific case displayed in Figure 9b,[256] and further elaborated to account for the low-frequency nonlinearity by Makarovic *et al.*[20]. It was assumed that the conducting domain walls in the [100]-oriented grain are parallel to the applied electric field, allowing the charges to flow along the domain wall, making a leakage path through the grain. As a consequence, the internal field in the [100]-oriented grains will reduce at low frequencies due to the leakage along the conductive walls, whereas the internal field in the [111]-oriented grains will be enhanced instead (actual calculated internal fields in the two grains as a function of driving frequency are shown in Figure 9c). It was argued that in grains where the electric field is enhanced, irreversible domain wall movements are promoted, which increases the piezoelectric response and leads to the unusual field-dependent increase of $d_{33}$ in the low frequency regime, as shown in Figure 9a.

**Figure 9.** Enhanced piezoelectric response due to conducting ferroelectric domain walls. a) Frequency dependence of the small signal piezoelectric $d_{33}$ coefficient of a polycrystalline BiFeO$_3$ ceramic at two different driving electric-field amplitudes.[67] The lines represent a guide to the eye. The arrow indicates the effect of increasing electric field on the low-frequency piezoelectric response. The inset shows a cAFM image confirming presence of conducting domain walls in the BiFeO$_3$ ceramics. (Reproduced with permission.[67, 257] Copyright 2015, Wiley-VCH.) b) Schematic illustration highlighting the domain wall orientation in [100]- and [111]-oriented grains with respect to the electric field axis, $E_{app}$, used as the basis of Maxwell-Wagner modeling.[20, 256] Because of the different orientation of conducting domain walls in the two grains, the local electric field, $E_{loc}$, is either reduced (see $E_{100}$) or enhanced (see $E_{111}$) at low frequencies due to conduction along domain walls oriented parallel to the electric field $E_{app}$ (which occurs more extensively in [100]-oriented grains due to the vertical orientation of domain walls). The enhanced electric field in the [100]-oriented grain results in enhanced domain wall contributions to the piezoelectric response, schematically indicated by the blue arrows and bands. c) Calculated change in $E_{loc}$ caused by the conducting domain walls in [100]- and [111]-oriented grains.[256]

In summary, the two selected examples demonstrate that both insulating and conducting domain walls allow for controlling the AC behavior at the bulk level. A particular focus of future studies could be on tuning the time constant of the heterogeneous systems via control of domain wall conductivity, which should allow to determine the usable frequency range where dielectric or piezoelectric properties are enhanced. Together with the improved understanding of the electronic transport properties at domain walls and their tunability, e.g., via point defects[13, 14, 23, 258] and polarization charges,[29] new pathways are now open for the development of next-generation capacitors, actuators, or sensors, providing an exciting playground for future studies on functional ferroelectric materials.

## 5. Conclusion and future perspective

The DC electronic transport behavior at ferroelectric domain walls has been studied intensively over the last decade and a fundamental framework has been developed, explaining their unusual physical properties. In contrast, much less is known about the AC response of the walls. Enabled by the progress in advanced imaging and nanoscale impedance spectroscopy techniques, however, investigations are now well on the way. Frequency-dependent studies have shown the existence of two characteristic regimes that will be of interest with respect to the development of future domain-wall based AC technology. In the low-frequency regime ($f \lesssim$ 1 MHz), electrode–wall junctions can be utilized to control electronic signals. At higher frequencies ($f \gtrsim$ 1 MHz), a variety of physical phenomena has been identified that can be exploited to tune the conductivity of the domain walls. For example, domain-wall vibrations around equilibrium position and electrical losses related to oscillating charge carriers, as well as acoustic wave reflection have been demonstrated to co-determine the AC response of ferroelectric domain walls in the investigated high-frequency regime.

Going beyond the frequencies covered by the available microscopy studies, macroscopic dielectric measurements[259-262] suggested that phonon contributions govern the domain wall response in the THz regime.[263] The interaction of domain walls and phonons is interesting with respect to heat transport, offering the possibility to control energy flux based on ferroelectric domain walls.[55-57] At the local scale, related experiments may be realized by coupling terahertz pulses to a sharp metallic AFM tip, which allows for a spatial resolution of ~10 nm.[264, 265]

Aside from an expansion of the frequency range accessible in local probe experiments, the application of novel approaches may provide new insight into the AC response of ferroelectric domain walls. One particularly promising but so far unexplored possibility is to investigate the impedance characteristics of individual isolated domain walls in nano-structured samples, analogous to so-called bicrystals (see Figure 4c).[148] By performing measurements on samples containing individual ferroelectric domain walls, an unprecedented frequency range could be covered within a single experiment, enabling a systematic analysis of features in the dielectric spectra and their relation to, e.g., the domain wall orientation and charge state.

Concerning domain-wall based AC technology, first applications in devices for radiofrequency and as AC-to-DC converts at the nanoscale have been proposed.[13, 36] For example, the electrode–wall junction at neutral domain walls in hexagonal manganites was utilized to rectify electrical currents with operational frequencies up to 1 MHz. For state-of-the art diode and thin-film transistor materials, such as $MoS_2$, silicon, or carbon nanomaterials, a correlation between the charge carrier mobility and the operational frequency was established, as displayed in **Figure 10**.[266] Interestingly, the mobility of charge carriers responsible for domain wall conductivity was found to be in the same range as carbon nanomaterials (e.g. ≈670 cm²/(Vs) for tail-to-tail domain walls in h-ErMnO$_3$[267] or ≈2970 cm²/(Vs) for head-to-head domain walls in LiNbO$_3$[268]). Thus, ultra-small domain-wall based diodes with operational frequencies up to 10 GHz may be realized in the future by further improving experimental approaches, i.e., via downscaling the size of the contacting electrode. Up to now, the application-oriented AC studies on ferroelectric domain walls are still at a very early stage and have scratched only the tip of the iceberg. However, building up on the recent discoveries, more advanced domain-wall based AC technology may soon become feasible, including filters, full wave rectifiers, transmission lines or wave guides, adding exciting new perspectives to the field of domain wall nanoelectronics.

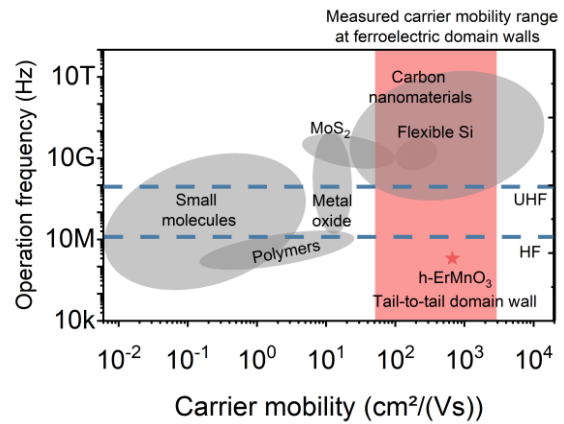

**Figure 10.** Operational frequency of diodes and thin-film transistors as a function of charge carrier mobility. Threshold frequencies for high frequency (HF, 13.56 MHz) and ultra-high frequency (UHF, 860 MHz) operation modes are displayed by the dashed lines. Data is taken from ref. [266] and references therein. The star indicates the operational frequency of a domain-wall based half-wave rectifiers using conducting tail-to-tail domain walls in h-ErMnO$_3$.[36] The red area indicates the expected mobility of charge carriers responsible for domain wall conductivity, based on measurements of h-ErMnO$_3$,[267] h-YbMnO$_3$,[269] and LiNbO$_3$[268].

In addition, the development of functional bulk ferroelectrics will strongly benefit from the ongoing advances in the characterization and understanding of the AC response at the domain walls. Intriguing examples are the utilization of domain wall networks as reconfigurable barrier layer capacitors[244] and the unusual low-frequency (sub-Hz) enhanced electromechanical properties in ferroelectric ceramics triggered by conducting walls[256]. In general, the possibility to create, move and erase ferroelectric domain walls on demand adds a new degree of freedom that can be utilized to tune dielectric, electronic, and electromechanical properties, paving the way towards a novel generation of materials for multifunctional AC devices. Going beyond single-phase ferroelectrics, heterostructures may be realized to combine the unusual AC conductivity of ferroelectric and magnetic domain walls,[53] enabling magnetic field control and novel magnetoelectric correlation phenomena, representing an exciting playground for frequency-dependent investigations on functional domain wall in the years to come.

**Acknowledgements**

The authors thank E. N. Lysne, D. M. Evans, and S. Krohns for fruitful discussions and for valuable input. J.S. acknowledges support from the Feodor Lynen Research Fellowship Program of the Alexander von Humboldt Foundation. T.R. thanks the Slovenian Research Agency for funding (research program P2-0105). D.M. thanks NTNU for support through the Onsager Fellowship Program, the Outstanding Academic Fellow Program, and acknowledges funding from the European Research Council (ERC) under the European Union's Horizon 2020 Research and Innovation Program (Grant Agreement No. 86691).